\documentclass[twocolumn, showpacs, prl,unsortedaddress]{revtex4}

\usepackage{graphicx}

\usepackage{dcolumn}

\usepackage{color}
\usepackage{bm}

\begin{document}

\title{Effective mass suppression upon complete spin-polarization in an isotropic two-dimensional electron system}

\date{\today}

\author{T.\ Gokmen}

\author{Medini\ Padmanabhan}

\author{K.\ Vakili}

\author{E.\ Tutuc}

\author{M.\ Shayegan}

\affiliation{Department of Electrical Engineering, Princeton
University, Princeton, NJ 08544}

\begin{abstract}

We measure the effective mass ($m^{*}$) of interacting
two-dimensional electrons confined to a 4.5 nm-wide AlAs quantum
well. The electrons in this well occupy a single out-of-plane
conduction band valley with an isotropic in-plane Fermi contour.
When the electrons are partially spin polarized, $m^{*}$ is larger
than its band value and increases as the density is reduced.
However, as the system is driven to full spin-polarization via the
application of a strong parallel magnetic field, $m^{*}$ is
suppressed down to values near or even below the band mass. Our
results are consistent with the previously reported measurements on
wide AlAs quantum wells where the electrons occupy an in-plane
valley with an anisotropic Fermi contour and effective mass, and
suggest that the effective mass suppression upon complete spin
polarization is a genuine property of interacting two-dimensional
electrons.

\end{abstract}


\maketitle

As the density of an interacting two-dimensional electron system
(2DES) is reduced, the interaction strength characterized by the
ratio $r_{s}$ of the Coulomb energy to Fermi energy, is enhanced.
In low disorder, dilute 2DESs the ground state properties are
dominated by the electron-electron interaction
\cite{CeperleyPRB1989}. In the Fermi liquid theory, interactions
modify the Fermi liquid parameters \cite{KwonPRB94} and
renormalize the effective mass ($m^*$) and the spin susceptibility
(${\chi}^*$ ${\propto}$ $g^{*} m^{*}$) of the 2DES, where $g^*$ is
the Lande g-factor. In particular, $\chi^*$ and $m^*$ are expected
to be larger than the band values ($\chi_b$ and $m_b$) for large
$r_{s}$ \cite{AttaccalitePRL2002, AsgariPRB2006, AsgariSSC2004,
DePaloPRL2005, GangadharaiahPRL05, ZhangPRL05, ZhangPRB05,
GokmenPRB07}. Indeed, enhancements of $\chi^*$ and $m^*$ at large
$r_{s}$ are observed in various 2DESs \cite{SmithPRL72, PanPRB99,
PudalovPRL02, ShashkinPRB02, ShashkinPRL03, VakiliPRL04,
TanPRL2005, FangPR1968, OkamotoPRL1999, VitkalovPRL2001,
ShashkinPRL2001, ZhuPRL2003, PrusPRB2003, TutucPRB2003,
TanPRB2006}. However, in 2DESs occupying wide AlAs quantum wells
(QWs) an unexpected trend is observed as the system becomes fully
spin polarized: $m^*$ is suppressed and falls to values near or
below $m_b$ even for $r_s$ values up to 21
\cite{PadmanabhanPRL08}. A subsequent study in similar samples
\cite{GokmenPRL08} revealed that the mass suppression disappears
when the electrons occupy two conduction-band valleys, signaling
that the mass suppression is unique to single-component (fully
spin and valley polarized) systems.

\begin{figure}
\centering
\includegraphics[scale=0.95]{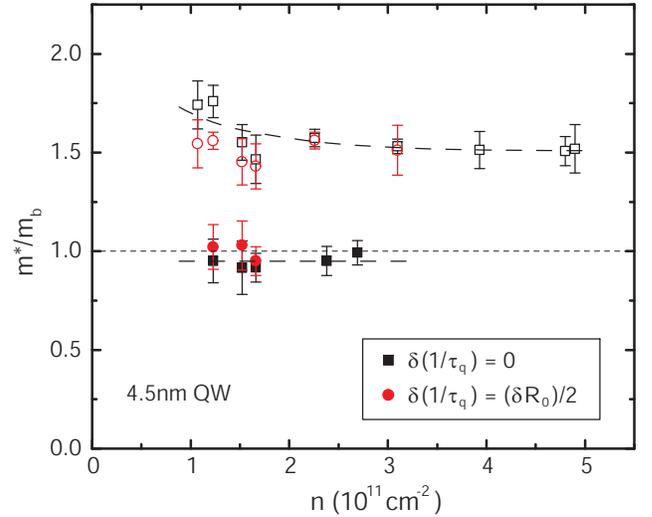}
\caption{(Color online) Effective mass, normalized to the band mass,
measured as a function of density for a 2DES confined to a 4.5
nm-wide AlAs quantum well. Open and closed symbols represent $m^*$
measured in partially and fully spin-polarized 2DESs, respectively.
Black and red points correspond to $m^*$ values deduced either
assuming a constant quantum lifetime $\tau_q$ or that the relative
temperature-dependence of $\tau_q$ is half the size of the relative
temperature-dependence of the background resistance, respectively.
Each data point represents $m^*$ averaged over different Landau
level filling factors, $\nu$, and the error bars include the
variation of $m^*$ with $\nu$. The dashed curves through the data
points are guides to the eye.}
\end{figure}

Here we report measurements of $m^*$ in the partially and fully spin
polarized regime as a function of density in a 2DES where the
electrons are confined to a 4.5 nm-wide AlAs QW. This 2DES is
different from the 2DESs used in Ref. \cite{PadmanabhanPRL08} in two
important aspects: it has a very $small$ layer thickness ($<$ 4.5
nm) and an $isotropic$ effective mass. Bulk AlAs has three
equivalent, ellipsoidal conduction band valleys at the X-points of
the Brillouin zone with longitudinal and transverse effective
masses, $m_l$=1.05 and $m_t$=0.205 (in units of the free electron
mass) \cite{ShayeganPSSb2006,MomosePhyE99,LayAPL93,Gunawan2007PRB}.
In samples of Ref. \cite{PadmanabhanPRL08}, the electrons are
confined to either an 11 nm- or 15 nm-wide AlAs QW and occupy one of
the two in-plane valleys with an anisotropic Fermi contour and
anisotropic band masses equal to 0.205 and 1.05, leading to $m_b$ =
$\sqrt{m_lm_t}$ = 0.46. In contrast, in the present, 4.5 nm-wide
AlAs QW, the electrons occupy a single, out-of-plane valley with an
isotropic Fermi contour and isotropic $m_b = m_t = 0.205$
\cite{mass}. In spite of these differences, our main findings,
summarized in Fig. 1, are consistent with the study in Ref.
\cite{PadmanabhanPRL08}: When the 2DES is partially spin-polarized
(open symbols), $m^*$ is larger than its band value and gradually
increases with decreasing density. But as we fully spin polarize the
2DES by subjecting it to strong parallel magnetic fields
\cite{BparNote}, $m^*$ is suppressed down to values near the band
mass (closed symbols). The two colors in Fig. 1 correspond to two
different types of analyses used for the $m^*$ determination, which
we will discuss later in the paper. Given that this system is close
to an ideal 2DES in the sense that it has a very small layer
thickness and an isotropic Fermi contour, it appears that mass
suppression upon full spin polarization is a genuine property of
interacting 2DESs.

We performed measurements on a sample grown on a GaAs (001)
substrate and consisting of a 4.5 nm-wide AlAs QW, flanked by
Al$_{0.4}$Ga$_{0.6}$As barriers \cite{VakiliPRL04,VakiliAPL2006,
ShayeganPSSb2006}. We patterned the sample in a Hall bar
configuration, and made ohmic contacts by depositing AuGeNi and
alloying in a reducing environment. Metallic front and back gates
were deposited to control the carrier density, $n$, which was
determined from the frequency of Shubnikov de Haas (SdH)
oscillations and from the Hall resistance. Values of $n$ in our
sample are in the range of 1.07 to 4.9 $\times$ 10$^{11}$
cm$^{-2}$, with mobilities $\mu$ = 1.4 to 4.9 m$^2$/Vs. Using the
AlAs dielectric constant of 10 and the band effective mass $m_b =
0.205$, our density range corresponds to 3.1 $<$ $r_s$ $<$ 6.7,
where $r_s$ is the ratio of the average inter-electron spacing
measured in units of the effective Bohr radius. The
magneto-resistance measurements were performed down to a
temperature ($T$) of 0.3 K, and up to a magnetic field of 31 T,
using low-frequency lock-in techniques. The sample was mounted on
a tilting stage to allow the angle, $\theta$, between the normal
to the sample and the magnetic field to be varied \textit{in
situ}.

\begin{figure} \centering
\includegraphics[scale=1]{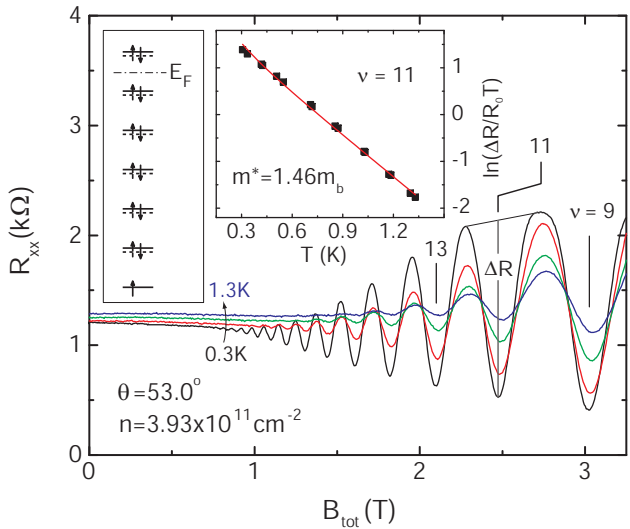}
\caption{(Color online) Magneto-resistance traces at a density of
3.93 $\times$ 10$^{11}$ cm$^{-2}$ and $\theta=53.0^{\circ}$. The
traces were taken at $T \cong$ 0.34, 0.71, 1.03 and 1.30 K. Insets
show the energy level diagram at this tilt angle (left) and the
Dingle fit at $\nu$ = 11 assuming a constant $\tau_q$ and $R_{o}$
(right).}
\end{figure}

\begin{figure} \centering
\includegraphics[scale=1]{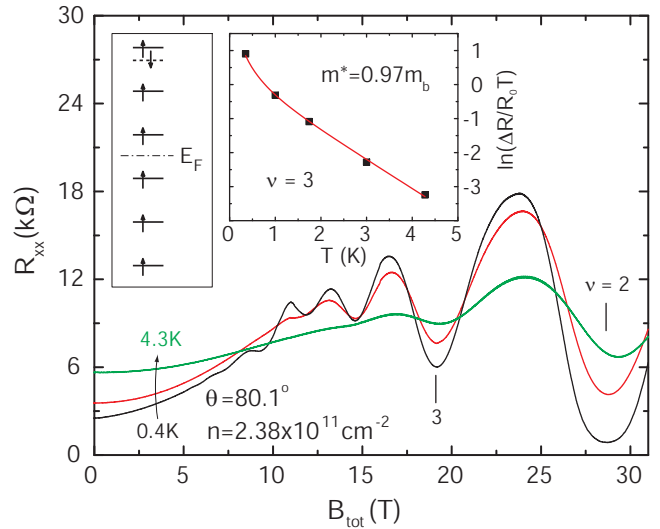}
\caption{(Color online) Magneto-resistance traces at a density of
2.38 $\times$ 10$^{11}$ cm$^{-2}$ and $\theta=80.1^{\circ}$. The
traces were taken at $T \cong$ 0.35, 1.74 and 4.28 K. Insets show
the energy level diagram at this tilt angle (left) and the Dingle
fit at $\nu$ = 3 assuming a constant $\tau_q$ and $R_{o}$ (right).}
\end{figure}

To deduce $m^{*}$, we analyzed the $T$-dependence of the strength
($\Delta R$) of the SdH oscillations using the standard Dingle
expression \cite{DinglePRSL52}: $\Delta
R/R_{o}=8exp(-\pi/\omega_{c}\tau_{q})\xi/sinh(\xi)$, where the
factor $\xi/sinh(\xi)$ represents the \textit{T}-induced damping
($\xi=2\pi^{2}k_{B}T/\hbar\omega_{c}$), and
\textit{$\omega_{c}=eB_{\bot}/m^{*}$} is the cyclotron frequency,
$B_{\bot}$ is the perpendicular component of the magnetic field,
$R_{o}$ is the non-oscillatory component of the resistance near a
SdH oscillation, and $\tau_{q}$ is the single-particle (quantum)
lifetime. We analyzed our data using two methods, each based on a
different set of assumptions. First, we assumed that both $R_{o}$
and $\tau_{q}$ are $T$-independent. This is the usual assumption,
commonly made when the $T$-dependence of $R_{o}$ is small. For our
sample the $T$-dependence of $R_{o}$ is indeed weak at high
densities (see, e.g., Figs. 2-3). At low densities, however, $R_{o}$
is $T$-dependent and, for the lowest densities, $R_{o}$ changes by
as much as 60\% in the temperature range of our data (see, e.g.,
Figs. 4 and 5), implying that $\tau_{q}$ can depend on $T$.
According to a theoretical study \cite{AdamovPRB2006}, for
short-range scatterers, the relative $T$-dependent correction to
$\tau_q$ is half of the relative correction to the transport
scattering time $\tau_{tr} \propto 1/R_{o}$. For long-range
scatterers, the $T$-dependent correction to $\tau_q$ is expected to
be smaller \cite{DasSarmaPrivate}. In our second analysis method, we
included the $T$-dependence of $R_{o}$, and assumed that the
relative $T$-dependence of $\tau_{q}$ is equal to half the relative
$T$-dependence of $R_{o}$ \cite{DingleExpNote}. Note that these two
methods should bound the maximum error in $m^*$ determination
introduced by the $T$-dependence of $\tau_{q}$
\cite{BackgroundResNote}.

Figure 2 shows representative data for the partially spin polarized
case at a relatively high density, $n$ = 3.93 $\times$ 10$^{11}$
cm$^{-2}$. The angle $\theta$ is set carefully so that the opposite
spin levels are at coincidence as shown in the left inset of Fig. 2
\cite{MassVakili}. Consistent with this energy level diagram, in the
magneto-resistance traces shown in Fig. 2, resistance minima at odd
$\nu$ are strong while the minima at even $\nu$ are entirely absent.
By fitting the amplitude of the SdH oscillations near $\nu=11$ to
the Dingle expression and assuming $T$-independent $\tau_{q}$ and
$R_{o}$, we obtain $m^{*}$ = 1.46 $m_{b}$ (see Fig. 2 right inset).
Moreover, as illustrated in the Dingle plot in Fig. 6(a), in the
whole temperature and magnetic field range the data set can be fit
to the Dingle expression by assuming two constants $\tau_{q}$ and
$m^*$. Since the background resistance in this case has a very small
temperature dependence, our second analysis method that assumes
$T$-dependent $\tau_{q}$ and $R_{o}$ yields essentially the same
$m^{*}$.

\begin{figure} \centering
\includegraphics[scale=1]{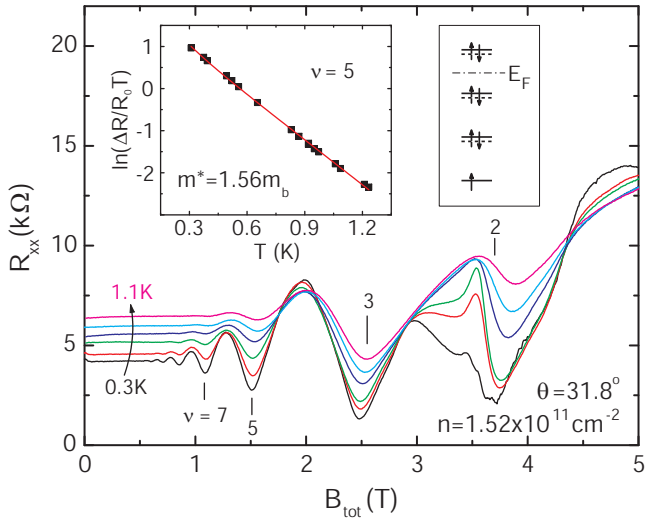}
\caption{(Color online) Magneto-resistance traces at a density of
1.52 $\times$ 10$^{11}$ cm$^{-2}$ and $\theta=31.8^{\circ}$. The
traces were taken at $T \cong$ 0.31, 0.49, 0.65, 0.83, 0.95 and 1.09
K. Insets show the energy level diagram at this tilt angle (right)
and the Dingle fit at $\nu$ = 5 using a constant $\tau_q$ and
$R_{o}$ (left).}
\end{figure}

In Fig. 3 we show data at the density of 2.38 $\times$ 10$^{11}$
cm$^{-2}$, at a very high tilt angle, $\theta=80.1^{\circ}$. At
this $\theta$, the magneto-resistance traces initially show a rise
with magnetic field because of the loss of screening with
increasing spin polarization \cite{LossofScreening}. The 2DES
becomes fully spin polarized above $B_{tot} \cong$ 11 T, and the
resistance minima at 2 $\leq \nu \leq$ 5 are clearly observed.
Note that at this angle, the lowest five Landau levels are spin
polarized as indicated by the energy level diagram shown in Fig. 3
left inset. To measure the fully spin polarized $m^*$ we fit the
amplitude of the SdH oscillations near $\nu=3$ to the Dingle
expression by assuming $T$-independent $\tau_{q}$ and $R_{o}$, and
we deduce $m^{*}$ = 0.97 $m_{b}$ (see Fig. 3 right inset). As
 is apparent from the magneto-resistance traces, although at zero
field there is a considerable change in resistance with
temperature, the background resistance at the SdH oscillation near
$\nu=3$ is small. Therefore including the $T$-dependence of
$R_{o}$ in the analysis does not make much difference and our
first and second analysis methods give essentially identical
results.

Now we present data at a lower density ($n$ = 1.52 $\times$
10$^{11}$ cm$^{-2}$) where temperature dependence of the background
is strong. Figure 4 shows data for the partially spin polarized case
for this density. Consistent with the energy level diagram in the
right inset of Fig. 4, $\theta$ is set to the coincidence angle so
that the resistance minima at odd $\nu$ are strong while the minima
at even $\nu$ are either entirely absent or are accompanied by a
spike (e.g., at $\nu=2$) \cite{DePoortereScience2000}. By fitting
the amplitude of the SdH oscillations near $\nu=5$ to the Dingle
expression and assuming $T$-independent $\tau_{q}$ and $R_{o}$, we
obtain $m^{*}$ = 1.56 $m_{b}$ (see Fig. 4 left inset). The Dingle
plot for this data set is also shown in Fig. 6(b). It is apparent
from the quality of the fit that single $\tau_{q}$ and $m^{*}$ can
explain the whole data set in the given temperature and magnetic
field range. However, as discussed before, the quality of the fit
does not justify the assumption of $\tau_{q}$ being $T$-independent.
In addition, as can be seen from the magneto-resistance traces in
Fig. 4, $R_{o}$ changes with $T$ as much as 50\% in the indicated
temperature range, implying that $\tau_{q}$ can also be
$T$-dependent. Therefore, applying our second analysis method, i.e.,
including the $T$-dependence of $R_{o}$ and assuming a $T$-dependent
$\tau_{q}$ that changes by 25\% in the same temperature range, we
deduce $m^{*}$ = 1.44 $m_{b}$.

\begin{figure} \centering
\includegraphics[scale=1]{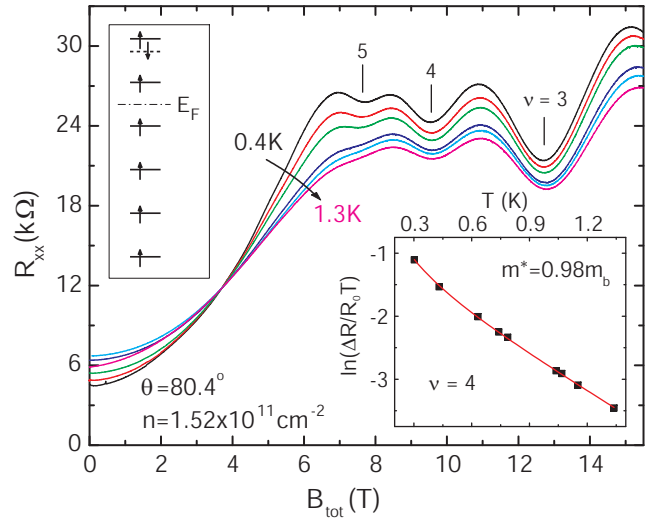}
\caption{(Color online) Magneto-resistance traces for the same
density as in Fig. 4 at $\theta=80.4^{\circ}$. The traces were taken
at $T \cong$ 0.43, 0.63, 0.78, 1.07, 1.15 and 1.30 K. Insets show
the energy level diagram for this tilt angle (left) and the Dingle
fit at $\nu$ = 4 using a constant $\tau_q$ and $R_{o}$ (right).}
\end{figure}

In Fig. 5 we show data at the same density as in Fig. 4 but at a
very high tilt angle, $\theta=80.4^{\circ}$. Main features of the
data are the same as in Fig. 3: Magneto-resistance traces show an
initial rise with magnetic field and the 2DES becomes fully spin
polarized above $B_{tot} \cong$ 7 T. Filling factors $\nu \leq$ 5
are in the fully spin polarized regime as indicated by the energy
level diagram shown in Fig. 5 left inset. To measure the fully
spin polarized $m^*$ we fit the amplitude of the SdH oscillations
near $\nu=4$ to the Dingle expression by assuming $T$-independent
$\tau_{q}$ and $R_{o}$, and we deduce $m^{*}$ = 0.98 $m_{b}$ (see
Fig. 3 right inset). As can be seen from the magneto-resistance
traces, the 2DES goes through a metal-insulator transition at
$B_{tot} \cong$ 3.7 T before the electrons become fully spin
polarized \cite{VakiliPRL04}. The background resistance around
$\nu=4$ therefore has an insulating behavior. Again using our
second method, i.e., including the $T$-dependence of $R_{o}$ and
assuming a $T$-dependent $\tau_{q}$ that is half as large as the
$T$-dependence of $R_{o}$, we deduce $m^{*}$ = 1.11 $m_{b}$.

\begin{figure} \centering
\includegraphics[scale=0.85]{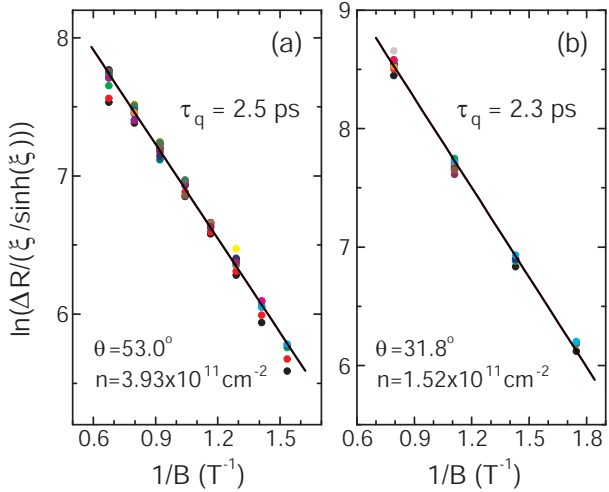}
\caption{(Color online) Dingle plots of $\Delta R/(\xi/sinh(\xi))$
vs. $1/B$ summarizing data taken for the partially spin polarized
case for: (a) a density of $3.93 \times 10^{11}$ cm$^{-2}$ in the
range $0.3 \lesssim T \lesssim 1.3$ K and $11 \leq \nu \leq 25$; and
(b) a density of $1.52 \times 10^{11}$ cm$^{-2}$ in the range $0.3
\lesssim T \lesssim 1.1$ K and $5 \leq \nu \leq 11$.}
\end{figure}

We analyzed data at various $\nu$ at several densities. Our results
are summarized in Fig. 1, where each data point represents $m^*$
averaged over different $\nu$, and the error bar includes the
variation of $m^*$ with $\nu$. The results from the first and second
analysis methods are shown in black and red in Fig. 1, respectively.
At high densities where the background is $T$-independent, the two
methods yield essentially identical results. However, as the density
of the 2DES is lowered, the $T$-dependent background becomes
stronger and the two methods give slightly different masses.
Independently of the method we use, our conclusions remain the same:
In the partially spin polarized case \cite{PolarizationNote} $m^*$
is enhanced over $m_b$ and increases with decreasing density, while
for the fully spin polarized system $m^*$ values are clearly
suppressed compared to the partially spin polarized case and are
very close to $m_b$.

\begin{figure} \centering
\includegraphics[scale=0.95]{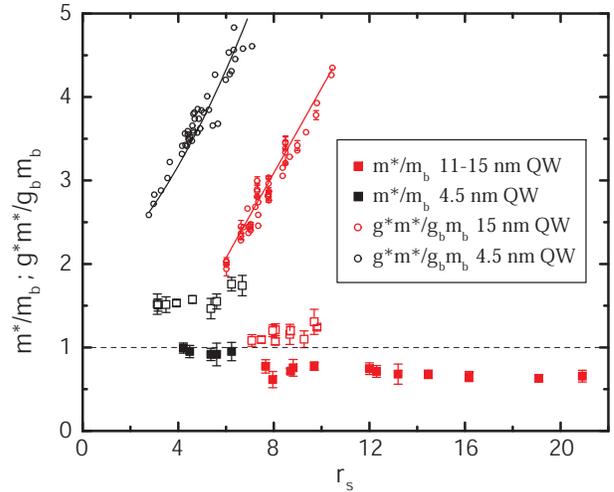}
\caption{(Color online) Normalized effective mass and spin
susceptibilities of both narrow and wide AlAs QWs as a function of
the interaction strength, $r_s$. Black and red circles are
$g^*m^*/g_bm_b$ of 4.5 nm and 15 nm-wide AlAs QWs taken from Refs.
\cite{VakiliPRL04} and \cite{GokmenPRB07}, respectively. Open and
closed red squares are $m^*/m_b$ for partially and fully spin
polarized system, respectively, for 11 to 15 nm-wide AlAs QWs from
Ref. \cite{PadmanabhanPRL08}. Black squares are $m^*/m_b$ data
measured in our sample for a 4.5 nm-wide AlAs QW.}
\end{figure}

As another summary plot, in Fig. 7 we show the normalized effective
mass and the spin susceptibilities of both narrow and wide AlAs QWs
as a function of interaction strength, $r_s$. Black and red circles
are the normalized spin susceptibilities of 4.5 nm and 15 nm-wide
AlAs QWs, taken from Refs. \cite{VakiliPRL04} and
\cite{GokmenPRB07}, respectively. Open and closed red squares
represent $m^*/m_b$ for partially and fully spin polarized system,
respectively, for the wide AlAs QWs of Ref. \cite{PadmanabhanPRL08}.
Black squares are the measured $m^*$ data for our sample that are
shown in Fig. 1.

In an ideal 2DES, the normalized values of the spin susceptibility
and effective mass each should follow a universal curve as a
function of $r_s$ \cite{AttaccalitePRL2002,AsgariPRB2006}. However
non-ideal factors such as finite layer thickness, anisotropy of the
Fermi contour, and disorder give non-universal corrections. Although
for high quality samples the effect of disorder is small
\cite{DePaloPRL2005,AsgariSSC2004}, finite layer thickness and
anisotropy of the Fermi contour modify the interaction significantly
\cite{DePaloPRL2005,GokmenPRB07,ZhangPRB05,Nonidealeffects} and
change the spin susceptibility and the effective mass
renormalization. Because of the small layer thickness and isotropic
Fermi contour of the electrons in narrow AlAs samples, the spin
susceptibility follows very closely the predictions of quantum Monte
Carlo calculations for an ideal 2D system (not shown)
\cite{AttaccalitePRL2002,VakiliPRL04,ShayeganPSSb2006,DePaloPRL2005}.
However, for wide AlAs samples the measured susceptibilities are
considerably lower than narrow AlAs samples because the strength of
the Coulomb interaction is reduced by the finite layer thickness
effect \cite{DePaloPRL2005,ZhangPRB05} and the anisotropy of the
Fermi contour \cite{GokmenPRB07}. It is clear from the data that the
partially spin polarized masses for wide AlAs QWs are also smaller
than for narrow AlAs QWs even though the $r_s$ values are larger.
Similar to the spin susceptibility case \cite{GokmenPRB07}, the
corrections to $m^*$ due to the layer thickness and the anisotropic
effective mass are expected to give smaller $m^*$ values in the
partially spin polarized case \cite{AsgariPRB2006, ZhangPRB05},
consistent with our data. In addition, we emphasize that the
mobility of the electrons in narrow quantum wells are much lower
compared to wide quantum wells because of the prevalence of the
interface roughness scattering. Therefore, it is also possible that
the higher disorder in narrow quantum wells is responsible for $m^*$
being larger \cite{AsgariSSC2004}. We point out that in the
partially polarized case there are also some quantitative
differences between our results on narrow AlAs QWs and the previous
studies done on Si-MOSFETs \cite{PudalovPRL02, ShashkinPRB02,
ShashkinPRL03} and GaAs 2DESs \cite{TanPRL2005}. It has been
reported that in GaAs 2DESs $m^*$ has a strong $r_s$ dependence
although $m^*$ values are much smaller compared to narrow AlAs QWs
for the similar $r_s$ range. It is likely that this discrepancy is
because of the larger finite layer thickness and less disorder in
2DESs in GaAs samples. On the other hand, because of the valley
degeneracy of Si-MOSFET samples, such a comparison is not valid: as
shown in Ref. \cite{GokmenPRL08}, the valley degeneracy affects the
mass renormalization considerably.

In the fully polarized regime, it is natural to also expect some
dependence of $m^*$ on the layer thickness, Fermi contour anisotropy
and disorder. However, it is not clear from the data whether $m^*$
are lower for wide AlAs samples because of non-ideal factors or
because these masses are measured at larger $r_s$ values. We
conclude that the mass suppression is very robust and is observed in
a very wide range of $r_s$ values and independent of sample and
system parameters such as disorder, layer thickness and anisotropy.

It is intuitively clear that, the spin polarization of the 2DES
should affect the $m^*$ re-normalization since it modifies the
exchange interaction. Naively, one might think that for a fully spin
polarized system the Fermi energy is doubled so the interaction is
weaker compared to the spin unpolarized case, and hence the mass for
the spin polarized case would be smaller. Although this argument
gives the correct qualitative behavior of $m^*$ for a fixed density,
it does not explain why $m^*$ for the fully polarized system stays
small (near or below $m_b$) even at very high $r_s$ values. Recent
theoretical work \cite{ZhangPRL05,GangadharaiahPRL05} has addressed
the role of spin-polarization on $m^*$ re-normalization. It is
reported in Ref. \cite{GangadharaiahPRL05} that $m^*$ very weakly
depends on the spin polarization for valley degenerate systems.
Since in our case the 2D electrons occupy a single valley, this is
not relevant to our data. The more relevant study \cite{ZhangPRL05},
which deals with a single-valley system, reports a rather strong
dependence of $m^*$ on the degree of spin-polarization. Although it
is predicted in Ref. \cite{ZhangPRL05} that for a fully
spin-polarized 2DES $m^*$ is smaller compared to the spin
unpolarized case, there remains major qualitative discrepancies with
our data. For example, $m^*$ for a fully spin polarized system is
predicted to increase with increasing $r_s$ and become smaller than
$m_b$ only for $r_s$ $<$ 2. In contrast, our data suggest that $m^*$
stays very close to $m_b$ even in the range 4 $<$ $r_s$ $<$ 6.
Including the data from Ref. \cite{PadmanabhanPRL08}, which extends
up to $r_s \simeq 21$, the discrepancy becomes even more serious. An
understanding of the magnitude and density dependence of $m^*$ for a
single component (single valley and fully spin polarized) 2DES
awaits future theoretical developments.

In summary, we confirmed the observation of $m^*$ suppression upon
full spin polarization in a system with a very small layer
thickness and isotropic Fermi contour. Since this system is very
close to an ideal 2DES, our data suggest that mass suppression for
a single component system is a general property of an interacting
2DES.

We thank the NSF for support. Part of this work was done at the
NHMFL, Tallahassee, which is also supported by the NSF. We thank E.
Palm, T. Murphy, B. Pullum and S. Maier for assistance.

\break


\begin{references}
\bibitem{DingleExpNote}
We emphasize that in all cases our data can be fit reasonably well
to the Dingle expression in the whole temperature and magnetic field
range by assuming two constants $\tau_{q}$ and $m^*$ (see, e.g.,
Fig. 6). However, this does not justify that $\tau_{q}$ is
independent of $T$; indeed a given data set can be fit equally well
to the Dingle expression with a different, fixed $m^*$, and a
$\tau_{q}$ that has some $T$-dependence. Since $m^*$ and the
$T$-dependence of $\tau_{q}$ cannot be determined independently, in
our second analysis method we use the $T$-dependence of the
background resistance to estimate the $T$-dependence of $1/\tau_{q}$
and deduce $m^*$.
\bibitem{BackgroundResNote}
Although the $T$-dependence of $R_{o}$ and $\tau_{q}$ tend to cancel
each other in the fitting of $m^*$, it is the $T$-dependence of
$\tau_{q}$ that dominates since $\tau_{q}$ appears inside the
exponential term.
\bibitem{MassVakili}
The partially spin polarized masses are also reported in Ref.
\cite{VakiliPRL04}. However those masses are not deduced at the
coincidence condition. Setting the coincidence condition is
important because when the spin-up and spin-down levels do not
overlap, the gap is smaller than the cyclotron energy and one may
get artificially larger values for $m^*$.
\bibitem{PolarizationNote}
In our study, the "partially-spin-polarized" regime corresponds to
spin polarization ranging from 4\% to 20\%.
\bibitem{Nonidealeffects}
In 2D systems the nonuniversal correction arising from the finite
layer thickness is a conceptually simple form-factor that modifies
the bare Coulomb interaction. Similarly, the Fermi contour
anisotropy can be incorporated in Coulomb interaction by making a
coordinate transformation that makes the Fermi surface isotropic but
the interactions anisotropic.
\end{references}
\end{document}